\documentclass[12pt]{article}

\usepackage{amsfonts}
\usepackage{graphicx}
\usepackage{mathrsfs}
\usepackage{amsmath}
\usepackage{amssymb}
\usepackage{float}
\usepackage{natbib}
\usepackage{fullpage}

\newtheorem{pro}{Proposition}
\newtheorem{thm}{Theorem}

\newcommand{\bm}{\boldsymbol}

\def\tr{\mathrm {tr}}

\def\B{{\bf B}}
\def\D{{\bf D}}
\def\R{{\bf R}}

\def\S{{\bf S}}
\def\I{{\bf I}}
\def\bmu{{\bm \mu}}

\def\X{{\bm X}}

\def\V{\bm V}
\def\x{{\bf x}}

\def\tr{\mathrm {tr}}

\def\bmv{{\bm\varepsilon}}
\def\bth{{\bm\mu}}
\def\bms{{\bm\Sigma}}
\def\diag{\mathrm {diag}}

\def\cp{\mathop{\rightarrow}\limits^{p}}
\def\cd{\overset{\mathcal{L}}{\longrightarrow}}

\begin{document}

\title{High Dimensional Spatial Rank Test for Two-Sample Location Problem}



\author{Long Feng \\
              Northeast Normal University\\
              {flnankai@126.com}           
}

\date{}

\maketitle

\begin{abstract}
This article concerns tests for the two-sample location problem when
the dimension is larger than the sample size. The traditional
multivariate-rank-based procedures cannot be used in high
dimensional settings because the sample scatter matrix is not
available. We propose a novel high-dimensional spatial rank test in
this article. The asymptotic normality is established. We can allow
the dimension being almost the exponential rate of the sample sizes.
Simulations demonstrate that it is very robust and efficient in a
wide range of distributions.
\end{abstract}

{\it keywords}{High dimensional Tests; Spatial sign; Spatial
rank}

\section{Introduction}
Nowadays, high-dimensional data have been generated in many areas,
such as microarray analysis, hyperspectral imagery. The traditional
statistical methods, which assume the dimension is fixed, may not
work in the high-dimensional settings. In the last decades,
statistican devoted many new methods to deal with high dimensional
data. Specially, many efforts have been devoted to high dimensional
hypothesis testing problems. See \cite{r7}, \cite{r1}, \cite{r20},
\cite{r5},  \cite{r4}, \cite{r2}, and \cite{r71} for two-sample
tests for means, \cite{r16}, \cite{r19}, \cite{r6} and \cite{r26}
for testing a specific covariance structure, \cite{r11}, \cite{r24}
and \cite{r9} for high-dimensional regression coefficients.

In this paper, we consider the high dimensional two sample location
problem. \cite{r1} proposed a test statistic by replacing the
Mahalanobis norm in Hotelling's $T^2$ test statistic with Euclidian
norm. To allow simultaneous testing for ultra-high dimensional data,
\cite{r5} proposed a test statistic by removing the square term in
\cite{r1}. However, both these two test statistics are not invariant
under scalar transformation, $\X\to \B \X$ where $\B$ is a diagonal
matrix. Recently, many scalar-invariant test statistics have been
constructed, such as \cite{r21}, \cite{r22}, \cite{r18}, \cite{r10},
\cite{r12}. All these methods are based on the multivariate normal
assumption or the diverging factor model \citep{r1}. These
assumptions are a little restrictive for application. For example,
the multivariate $t$-distribution or mixture of multivariate normal
distribution do not belong to them. Moreover, the performance of
these moment based tests would be degraded for heavy-tailed
distributions.

In the traditional fixed dimension circumstance, multivariate sign
or rank based methods are often used to construct robust tests.
Those test statistics are very efficient and distribution free under
mild assumptions, or asymptotic so. However, those classic
multivariate sign or rank based tests also cannot be directly used
in high dimensional data because the scatter matrix is not
available. Recently, \cite{r23} proposed a high-dimensional
nonparametric multivariate test for one sample location problem
based on spatial-signs. \cite{r27} also proposed a high-dimensional
spatial sign test for the one sample problem by restricting to
spherical cases.  \cite{r8} proposed a scalar invariant test
statistic based on spatial sign for two sample location problem.
They show that their proposed test is robust and efficient for a
wide range of distributions. However, to estimate the location
parameter, they only can allow the dimension being the square of the
sample sizes at most. It is too restrictive for application. Thus,
we need to propose a new robust test procedure to allow for
ultra-high dimensional data.

Multivariate rank based methods also perform very efficient in
constructing robust test procedures. In this article, we propose a
high-dimensional spatial rank test for two sample location problem.
First, we estimate the scale of each variable by spatial rank based
procedures. Then, we propose our high dimensional spatial-rank test
based on leave out method. The test statistic is scalar-invariant
and treats all the variables in a ``fair" way. Furthermore, unlike
the spatial sign, we do not need to estimate the location parameters
for spatial rank. Thus, there are no bias term in our test statistic
when the dimension increases. So we can allow the dimension being
almost exponential rate of the sample sizes. We also establish its
asymptotic normality and propose the asymptotic relative efficiency
with respect to \cite{r18}'s test. Simulation studies show that our
test performs better than other moment-based test procedures under
heavy-tailed distributions. And when the dimension is ultra-high
against the sample sizes, our test would be more powerful than
\cite{r8}'s test because of its conservation. All the detailed
proofs are given in the Supplementary Material.

\section{High-Dimensional Spatial-rank test}
\subsection{The proposed test statistic}
Assume $\{\X_{i1},\cdots,\X_{in_i}\}$ for $i=1,2$ are two independently and identically distributed (i.i.d.) random
samples from $p$-variate elliptical distribution with density functions
\[\det(\bms_i)^{-1/2}g(||\bms_i^{-1/2}(\x-\bmu_i)||)\]
where $\bmu_i$'s are the symmetry centers and $\bms_i$ is the
positive definite symmetric $p\times p$ scatter matrix. Here we
consider the equal scatter matrix case, i.e. $\bms=\bms_1=\bms_2$.
We wish to test
\begin{align}\label{ht}
H_0:\bth_1=\bth_2\ \ \mbox{versus}\ \  H_1:\bth_1\neq\bth_2.
\end{align}
Hotelling's $T^2$ test statistic
$H_n=\frac{n_1n_2}{n}(\bar{\X}_1-\bar{\X}_2)^T\S_n^{-1}(\bar{\X}_1-\bar{\X}_2)$
is the classic method to deal with this two sample location problem,
where $n=n_1+n_2$, $\bar{\X}_i$ are the sample mean and $\S_n$ is
the pooled sample covariance matrtix. However, it is not very
efficient for the heavy-tailed distributions. Multivariate sign or
rank based methods are often used to construct robust test for these
location testing problem \citep{r17}. Define the spatial sign
function $U(\x)=\x/||\x||I(\x\not=\bm 0)$. When the dimension $p$ is
fixed, the spatial rank test using inner standardization is
\begin{align*}
Q^2=np\frac{\sum_{i=1}^2n_i||\tilde{\mathcal{V}}_i||^2}{\sum_{i=1}^2\sum_{j=1}^{n_i}||\hat{\mathcal{V}}_{ij}||^2},
\end{align*}
where
$\tilde{\mathcal{V}}_i=n_i^{-1}\sum_{j=1}^{n_i}\hat{\mathcal{V}}_{ij}$,
$\hat{\mathcal{V}}_{ij}=n_i^{-1}\sum_{k=1}^{n_i}U(\S^{-1/2}(\X_{ij}-\X_{ik}))$
and the full rank transformation matrix $\S^{-1/2}$ satisfy
\begin{align}\label{rv}
{\rm RCOV}\doteq
\frac{1}{n}\sum_{i=1}^2\sum_{j=1}^{n_i}\hat{\mathcal{V}}_{ij}\hat{\mathcal{V}}_{ij}\propto
\I_p.
\end{align} In traditional fixed $p$ circumstance, $Q^2$ is
affine-invariant and very robust. When the distribution is
heavy-tailed, $Q^2$ is  more efficient than the classic Hotelling's
$T^2$ test. However, when the dimension $p$ is larger than the
sample size $n$, the scatter matrix is not available and then $Q^2$
is not well defined.  Alternatively, we could estimate the diagonal
matrix of $\bms$ which is able to treat all the variables in a
``fair" way.

For the sample $\{\X_{ij}\}_{j=1}^{n_i}$, similar to (\ref{rv}), we
suggest to find a diagonal matrix $\mathcal{D}_i$ satisfy
\begin{align*}
\diag\left\{\frac{1}{n_i}\sum_{j=1}^{n_i}
R(\mathcal{D}_i^{-1/2}\X_{ij})R(\mathcal{D}_i^{-1/2}\X_{ij})\right\}
\propto \I_p,
\end{align*}
where
\[R(\mathcal{D}_i^{-1/2}\X_{ij})=\frac{1}{n_i}\sum_{k=1}^{n_i}U(\mathcal{D}_i^{-1/2}(\X_{ij}-\X_{ik})).\]
Thus, we adopt the following recursive algorithm
\begin{align*}
\mathcal{D}_i \leftarrow \mathcal{D}_i^{1/2}
\diag\left\{\frac{1}{n_i}\sum_{j=1}^{n_i}
R(\mathcal{D}_i^{-1/2}\X_{ij})R(\mathcal{D}_i^{-1/2}\X_{ij})\right\}
\mathcal{D}_i^{1/2},~~ \mathcal{D}_i \leftarrow
\frac{p}{\tr(\mathcal{D}_i)}\mathcal{D}_i.
\end{align*}
The resulting estimators of diagonal matrix are denoted as
$\hat{\D}_i$. We may use the sample variances as the initial
estimators. Note that we fixed $\tr(\mathcal{D}_i)=p$ in this
algorithm. Since $U(\sigma \x)=U(\x)$, without loss of generality,
we assume $\tr(\bms)=p$ in the following. So $\hat{\D}_i$ is a
consistent estimator of the diagonal matrix of $\bms$, i.e. $\D$
(Lemma \ref{le2} in the Appendix).

A natural idea is mimicking \cite{r5} and considering the following test statistic
\begin{align*}
G_{n}&=\frac{\sum_{i\not=j}^{n_1}\hat{\V}_{1i}^{T}\hat{\V}_{1j}}{n_1(n_1-1)}+\frac{\sum_{i\not=j}^{n_2}\hat{\V}_{2i}^{T}\hat{\V}_{2j}}{n_2(n_2-1)}
-2\frac{\sum_{i=1}^{n_1}\sum_{j=1}^{n_2}\hat{\V}_{1i}^{T}\hat{\V}_{2j}}{n_1n_2}.
\end{align*}
where
$\hat{\V}_{ij}=n^{-1}\sum_{k=1}^2\sum_{l=1}^{n_k}R(\hat{\D}^{-1/2}(\X_{kl}-\X_{ij}))$
is the  sample ``spatial-rank" of $\X_{ij}$ in the total samples.
And $\hat{\D}$ is also the estimator of $\D$ based on the total
samples. However, there are two drawbacks. First, there would be a
non-negligible bias term in $G_n$ with the growth of dimension
\citep{r8}. Second, those terms of
$R(\hat{\D}^{-1/2}(\X_{il}-\X_{ij}))$ are useless in detecting the
difference between these two samples. Thus, based on leave out
method and excluding $R(\hat{\D}^{-1/2}(\X_{il}-\X_{ij}))$ terms, we
propose the following high dimensional spatial rank test statistic
\begin{align*}
T_n=\frac{1}{n_1(n_1-1)}\frac{1}{n_2(n_2-1)}  \underset{i\not=j}{\sum^{n_1}\sum^{n_1}}
 \underset{s\not=t}{\sum^{n_2}\sum^{n_2}}U(\hat{\D}_{(i,j,s,t)}^{-1/2}(\X_{1i}-\X_{2s}))^TU(\hat{\D}_{(i,j,s,t)}^{-1/2}(\X_{1j}-\X_{2t})),
\end{align*}
where
$\hat{\D}_{(i,j,s,t)}=\frac{n_1}{n}\hat{\D}_{1(i,j)}+\frac{n_2}{n}\hat{\D}_{2(s,t)}$,
$\hat{\D}_{1(i,j)}$ and $\hat{\D}_{2(s,t)}$ are the corresponding
estimators of the diagonal matrix with leave-two-out samples
$\{\X_{1k}\}_{k\not=i,j}$, $\{\X_{2k}\}_{k\not=s,t}$, respectively.
Obviously, $U(\hat{\D}_{(i,j,s,t)}^{-1/2}(\X_{1i}-\X_{2s}))$ will be
deviate from zero if $\bmu_1\not=\bmu_2$ and then we will reject the
null hypothesis with large values of $T_n$. As shown later, the
expectation of $T_n$ is asymptotic negligible compared to its
standard deviation under $H_0$. We do not need a bias correction
procedure \citep{r10}. Moreover, the value of $T_n$ remains
unchanged for $\tilde{\X}_{ij}=\tilde{\D}\X_{ij}+\bm c$, where
$\tilde{\D}=\diag\{\tilde{d}_1^2,\cdots,\tilde{d}_p^2\}$,
$\tilde{d}_i$ are non-zero constants and $\bm c$ is a constant
vector. Our proposed test statistic is invariant under location
shift and the group of scalar transformations.

\subsection{Theoretical Results} We need the following conditions
for asymptotic analysis:  as $n, p\to \infty$,
\begin{itemize}
\item[(C1)] $n_1/n\to \kappa \in (0,1)$;
\item[(C2)] $\tr(\R^4)=o(\tr^2(\R^2))$ where $\R=\D^{-1/2}\bms\D^{-1/2}$;
\item[(C3)] $\log(p)=o(n)$ and $\tr(\R^2)-p=o(n^{-1}p^2)$;
\end{itemize}
Condition (C2) is the same as the condition (4) in \cite{r18}.
Condition (C3) is used to get the consistency of the diagonal matrix
estimators. To reduce the the difference between
$\D^{-1/2}(\X_{ij}-\bmu_i)$ and $\bm \varepsilon_{ij}\doteq
\bms^{-1/2}(\X_{ij}-\bmu_i)$, we require the correlation between
those variables to be not very strong and the dimension to be higher
enough.

First, we establish the asymptotic null distribution of $T_n$.
\begin{thm}
Under Conditions (C1)-(C3) and $H_0$, as $(p,n)\to \infty$,
\begin{align*}
{T_n}/{\sigma_n}\cd N(0,1),
\end{align*}
where $\sigma_n^2\doteq
\left(\frac{1}{2n_1(n_1-1)p^2}+\frac{1}{2n_2(n_2-1)p^2}+\frac{1}{n_1n_2p^2}\right)\tr(\R^2)$.
\end{thm}
In order to construct the test procedure, we still need to estimate
$\tr(\R^2)$. Here we proposed the following three ratio-consistent
estimator of $\tr(\R^2)$, $s=1,2$,
\begin{align*}
\widehat{\tr(\R^2)}_s=&\frac{2p^2}{P_{n_s}^4}\sum^*
U\left(\hat{\D}^{-1/2}_{s(i_1,i_2,i_3,i_4)}(\X_{si_1}-\X_{si_2})\right)^TU\left(\hat{\D}^{-1/2}_{s(i_1,i_2,i_3,i_4)}(\X_{si_3}-\X_{si_4})\right)\\
&\times U\left(\hat{\D}^{-1/2}_{s(i_1,i_2,i_3,i_4)}(\X_{si_3}-\X_{si_2})\right)^TU\left(\hat{\D}^{-1/2}_{s(i_1,i_2,i_3,i_4)}(\X_{si_1}-\X_{si_4})\right),\\
\widehat{\tr(\R^2)}_3=&\frac{p^2}{n_1^2n_2^2}\underset{i_1\not=i_2}{\sum^{n_1}\sum^{n_1}}
 \underset{i_3\not=i_4}{\sum^{n_2}\sum^{n_2}} \left(U\left(\hat{\D}^{-1/2}_{(i_1,i_2,i_3,i_4)}(\X_{1i_1}-\X_{1i_2})\right)^T
 U\left(\hat{\D}^{-1/2}_{(i_1,i_2,i_3,i_4)}(\X_{2i_3}-\X_{2i_4})\right)\right)^2.
\end{align*}
where $\hat{\D}_{s(i_1,i_2,i_3,i_4)}$ are the corresponding
estimators of diagonal matrix with leave-four-out samples
$\{\X_{sj}\}_{k\not=i_1,i_2,i_3,i_4}$. Throughout this article, we
use $\sum\limits^{*}$ to denote summations over distinct indexes.
For example, in $\widehat{\tr(\R^2)}_1$, the summation is over the
set $\{i_1\not=i_2\not=i_3\not=i_4\}$, for all
$i_1,i_2,i_3,i_4\in\{1,\cdots,n_1\}$ and $P_n^m={n!}/{(n-m)!}$.
\begin{pro}
Under Condition (C1)-(C3), as $(p,n)\to \infty$, we have
\[\frac{\widehat{\tr(\R^2)}_s}{\tr(\R^2)}\cp 1, ~~s=1,2,3.\]
\end{pro}
Here we use the following ratio-consistent estimator for
$\sigma_n^2$,
\[\hat{\sigma}_n^2=\frac{1}{2n_1(n_1-1)p^2}\widehat{\tr(\R^2)}_1+\frac{1}{2n_2(n_2-1)}\widehat{\tr(\R^2)}_2+\frac{1}{n_1n_2p^2}\widehat{\tr(\R^2)}_3.\]
This result suggests rejecting $H_0$ with $\alpha$ level of
significance if $T_n/\hat{\sigma}_n>z_{\alpha}$ where $z_{\alpha}$
is the upper $\alpha$ quantile of $N(0,1)$.

Next, we consider the asymptotic distribution of $T_n$ under the
alternative hypothesis
\begin{itemize}
\item[(C4)] $(\bmu_1-\bmu_2)^T\D^{-1}(\bmu_1-\bmu_2)=O(c_0^{-2}\sigma_n)$,
 $(\bmu_1-\bmu_2)^T\D^{-1/2}\R\D^{-1/2}(\bmu_1-\bmu_2)=o(npc_0^{-2}\sigma_n)$ where $c_0=E(||\D^{-1/2}(\X_{ij}-\X_{ik})||^{-1})$.
\end{itemize}
Condition (C4) require the difference between $\bmu_1$ and $\bmu_2$
is not large so that the variance of $T_n$ is still asymptotic
$\sigma_n^2$. And then we can propose the explicit power expression
of our test.

\begin{thm}
Under Conditions (C1)-(C4), as $(p,n)\to \infty$, we have
\[\frac{T_n-c_0^2(\bmu_1-\bmu_2)^T\D^{-1}(\bmu_1-\bmu_2)}{\sigma_n}\cd N(0,1).\]
\end{thm}

As a consequence, the asymptotic power of our proposed test
(abbreviated as SR hereafter) becomes
\begin{align*}
\beta_{\rm
SR}(||\bth_1-\bth_2||)&=\Phi\left(-z_{\alpha}+\frac{2c_0^2pn\kappa(1-\kappa)(\bth_1-\bth_2)^T\D^{-1}(\bth_1-\bth_2)}{\sqrt{2\tr({\R}^2)}}\right).
\end{align*}
In comparison, \cite{r18} show that the asymptotic power of their
proposed test (abbreviated as PA hereafter) is
\begin{align*}
\beta_{\rm
PA}(||\bth_1-\bth_2||)&=\Phi\left(-z_{\alpha}+\frac{np\kappa(1-\kappa)(\bth_1-\bth_2)^T{\D}^{-1}(\bth_1-\bth_2)}{E(||\bm
\varepsilon_{ij}||^2)\sqrt{2\tr({\R}^2)}}\right).
\end{align*}
The asymptotic relative efficiency (ARE) of SR with respect to PA is
\begin{align*}
{\rm ARE}({\rm SR}, {\rm
PA})=2c_0^2E(||\bm\varepsilon_{ij}||^2)\approx& 2\{E(||\bm\varepsilon_{1}-\bmv_2||^{-1})\}^2E(||\bm \varepsilon||^2)\\
=&\{E(||\bm\varepsilon_{1}-\bmv_2||^{-1})\}^2E(||\bm
\varepsilon_1-\bm \varepsilon_2||^2)\ge 1.
\end{align*}
by the Cauchy inequality and
$c_0=E(||\bm\varepsilon_{1}-\bmv_2||^{-1})(1+o(1))$ under Condition
(C4) (see the proof of Theorem 1). If $||\bm \varepsilon_1-\bm
\varepsilon_2||^2/E(||\bm \varepsilon_1-\bm \varepsilon_2||^2)\cp
1$, SR is equivalent to PA. Otherwise, our SR test would be more
efficient.

The ARE values for multivariate $t$-distribution with $\nu=3,4,5,6$
are  1.98, 1.48, 1.31, and 1.22, respectively. Clearly, the SR test
is more powerful than PA when the distributions are heavy-tailed
($\nu$ is small), which is also verified by simulation studies in
Section 3.

\section{Simulation}
Here we report a simulation study designed to evaluate the
performance of the proposed SR test. All the simulation results are
based on 2,500 replications. We consider the following five scenarios:
\begin{itemize}
\item[(I)] Multivariate normal distribution. $\X_{ij}\sim N(\bmu_i,\R)$.
\item[(II)] Multivariate normal distribution with different component variances.
 $\X_{ij}\sim N(\bmu_i,\bms)$, where $\bms=\D^{1/2}\R\D^{1/2}$ and $\D=\diag\{d_{1}^2,\cdots,d_{p}^2\}$, $d_{j}^2=3$, $j\le p/2$ and $d_{j}^2=1$, $j>p/2$.
\item[(III)] Multivariate $t$-distribution $t_{p,3}$.   $\X_{ij}$'s are generated from $t_{p,3}$ with $\bms=\R$.
\item[(IV)] Multivariate $t$-distribution with different component variances. $\X_{ij}$'s are generated from $t_{p,3}$ and $d_{j}^2$'s are generated from $\chi^2_2$.
\item[(V)] Multivariate mixture normal distribution $\mbox{MN}_{p,\gamma,9}$. $\X_{ij}$'s are generated from  $\gamma
f_p(\bmu_i,\R)+(1-\gamma)f_p(\bmu_i,9\R)$, denoted by
$\mbox{MN}_{p,\gamma,9}$, where $f_p(\cdot;\cdot)$ is the density
function of $p$-variate multivariate normal distribution. $\gamma$
is chosen to be 0.8.
\end{itemize}
First, we consider the low dimensional case $p < n$ and compare the
SR test with the traditional spatial-rank-based test $Q^2$
(abbreviated as TR). The common correlation matrix is
$\R=(0.5^{|j-k|})$.  For power comparison,  we consider the same
configurations of $H_1$:
$\eta=:||\D^{-1/2}(\bth_1-\bth_2)||^2/\sqrt{\tr(\R^2)}=0.5$. Without
loss of generality, under $H_1$, we fix $\bth_1=0$ and choose
$\bth_2$ as follows. The percentage of $\mu_{1l}=\mu_{2l}$ for
$l=1,\cdots,p$ are chosen to be $95\%$ (Sparse Case) and $50\%$
(Dense Case), respectively. At each percentage level, all the
nonzero $\mu_{2l}$ are equal. Two combinations of $(n_i,p)$ are
considered: $(30,24)$ and $(40,32)$. Table \ref{t1} reports the
empirical sizes and power of these two tests. The empirical sizes of
TR is significantly smaller than the nominal level. However, our SR
test can control the empirical sizes in most cases. In addition, our
SR test is more powerful than the TR test in all cases. This
findings are consistent with the results in \cite{r1}. Classical
Mahalanobis distance may lose efficiency because of the
contamination bias in estimating the covariance matrix with large
$p$. When $p/n\to c \in (0,1)$, having the inverse of the estimate
of the scatter matrix in constructing tests would be no longer
beneficial.

 \begin{table}
           \centering
           \caption{Empirical Size and  power comparison at 5\% significance when $p<n$.}
           \vspace{0.1cm}
      \renewcommand{\arraystretch}{1.2}
     \tabcolsep 5pt
         \begin{tabular}{ccccccccccccccc}\hline \hline
    &  \multicolumn{4}{c}{Size}    & & \multicolumn{4}{c}{Dense Case}&  &  \multicolumn{4}{c}{Sparse Case}     \\
$(n_i,p)$ & \multicolumn{2}{c}{(30,24)} & \multicolumn{2}{c}{(40,32)} & & \multicolumn{2}{c}{(30,24)} & \multicolumn{2}{c}{(40,32)} & & \multicolumn{2}{c}{(30,24)} & \multicolumn{2}{c}{(40,32)} \\
Scenario  & TR &SR & TR &SR &  & TR &SR & TR &SR & & TR &SR & TR &SR  \\
(I)  &1.6 &6.3 &1.3 &5.4  &  &27 &63 &41 &96  &  &42 &58 &63 &96\\
(II) &0.7 &6.5 &1.8 &5.2  &  &27 &64 &43 &98  &  &42 &58 &69 &96\\
(III)&1.4 &6.4 &0.8 &5.7  &  &19 &50 &31 &82  &  &29 &45 &46 &81 \\
(IV) &1.2 &4.9 &1.1 &5.3  &  &51 &60 &72 &91  &  &29 &54 &54 &59\\
(V)  &1.2 &5.3 &0.9 &4.5  &  &18 &45 &25 &78  &  &24 &44 &43 &76\\ \hline \hline
               \end{tabular}\label{t1}
           \end{table}

Next, we consider the high-dimensional cases, $p>n$, and compare the
SR with the tests proposed by \cite{r5} (abbreviated as CQ
hereafter), \cite{r18} and \cite{r8} (abbreviated as SS hereafter).
The sample size $n_i$ is chosen as $n_1=n_2=20$. Four dimensions
$p=100,200,400,800$ are considered. The other settings are all the
same as low dimensional cases. Table \ref{t3} reports the empirical
sizes and power of these four tests under normal and non-normal
cases. The sizes of CQ, PA and SR tests are generally close to the
nominal level under all the scenarios. In contrast, the sizes of the
SS test are a little smaller than 5\%, i.e., too conservative when
$p/n^2$ is large. It is not strange because SS test can only allow
the dimension $p$ being the square of the sample size $n$. As shown
in \cite{r8}, when $p/n^2$ is larger, there would be a
non-negligible bias term in SS test statistic because of the
estimation of location parameters. However, our SR test can allow
the dimension being almost the exponential rate of the sample sizes.
Consequently, our SR will be more powerful than the SS test when
$p/n^2$ is large. Moreover, under the normal cases (Scenarios (I)
and (II)), SS and SR tests perform similar to PA test. However,
under the non-normal cases (Scenarios (III)-(V)), both SS and SR
tests are clearly more efficient than PA and CQ tests. It is
consistent with the theoretical results in Section 2. Our SR test is
more robust and efficient when the distribution is heavy-tailed.
Finally, SS, PA and SR tests are more powerful than CQ test when the
variance of each variables are not equal (Scenario (II) and (IV)),
which demonstrates that a scalar-invariant test is needed.

 \begin{table}[ht]
           \centering
           \caption{Empirical Size and  power comparison at 5\% significance with equal scatter matrix}
           \vspace{0.1cm}
      \renewcommand{\arraystretch}{1.2}
     \tabcolsep 5pt
         \begin{tabular}{ccccccccccccccc}\hline \hline
    &  \multicolumn{4}{c}{Size}  &   & \multicolumn{4}{c}{Dense Case}&  &  \multicolumn{4}{c}{Sparse Case}     \\
          {$(n_i,p)$}                 &  \multicolumn{1}{c}{CQ}  & \multicolumn{1}{c}{SS} & \multicolumn{1}{c}{PA}   & \multicolumn{1}{c}{SR}&  &  \multicolumn{1}{c}{CQ}  & \multicolumn{1}{c}{SS} & \multicolumn{1}{c}{PA}   & \multicolumn{1}{c}{SR}& &  \multicolumn{1}{c}{CQ}  & \multicolumn{1}{c}{SS} & \multicolumn{1}{c}{PA}   & \multicolumn{1}{c}{SR}  \\\hline
          \multicolumn{15}{c}{Scenario (I)} \\
       (20,100) &6.9 &4.8 &5.0 &6.3 & &83  &77 &77 &82  & &79 &73 &74 &79 \\
       (20,200) &5.3 &3.4 &3.3 &5.0 & &87  &81 &83 &86  & &86 &81 &82 &87\\
       (20,400) &5.5 &2.5 &4.1 &4.7 & &90  &80 &85 &90  & &89 &80 &85 &90 \\
       (20,800) &5.4 &1.2 &4.0 &4.5 & &92  &81 &89 &92  & &91 &79 &88 &91 \\ \hline
                 \multicolumn{15}{c}{Scenario (II)} \\
       (20,100) &7.5 &5.0 &4.3 &6.0 & &44 &78 &78 &82 &  &40 &73 &73 &78\\
       (20,200) &4.5 &3.1 &2.5 &4.4 & &47 &81 &82 &87 &  &44 &81 &82 &87\\
       (20,400) &6.0 &2.5 &4.1 &5.7 & &46 &81 &86 &91 &  &42 &80 &85 &90\\
       (20,800) &6.6 &1.2 &3.8 &5.6 & &42 &80 &85 &90 &  &45 &79 &88 &91\\ \hline
%
          \multicolumn{15}{c}{Scenario (III)} \\
       (20,100)  & 5.8 & 4.3 & 3.3 & 6.2 & & 42 & 64 & 29 & 62 & & 37 & 59 & 29 & 58\\
       (20,200)  & 4.2 & 3.0 & 1.5 & 4.7 & & 44 & 67 & 34 & 69 & & 41 & 63 & 27 & 63\\
       (20,400)  & 6.1 & 2.0 & 3.9 & 5.7 & & 44 & 63 & 32 & 67 & & 41 & 63 & 31 & 66\\
       (20,800)  & 4.4 & 0.7 & 3.7 & 5.3 & & 44 & 61 & 32 & 72 & & 44 & 60 & 32 & 71\\ \hline
                 \multicolumn{15}{c}{Scenario (IV)} \\
       (20,100)  & 5.8 & 4.3 & 3.3 & 6.3 & & 11 & 68 & 33 & 67 & & 70 & 70 & 33 & 69\\
       (20,200)  & 7.5 & 3.0 & 1.7 & 4.4 & & 85 & 73 & 33 & 73 & & 13 & 65 & 28 & 65\\
       (20,400)  & 6.4 & 2.0 & 3.9 & 5.7 & & 10 & 64 & 30 & 69 & & 90 & 62 & 29 & 66\\
       (20,800)  & 5.7 & 0.7 & 3.7 & 5.2 & & 9.6& 60 & 32 & 73 & & 8.6& 62 & 32 & 72\\ \hline
                        \multicolumn{15}{c}{Scenario (IV)} \\
       (20,100)  & 7.5 & 4.0 & 5.5 & 5.2 & & 42 & 59 & 35 & 60 & & 35 & 55 & 30 & 57 \\
       (20,200)  & 6.5 & 3.0 & 4.5 & 5.2 & & 43 & 60 & 32 & 63 & & 41 & 56 & 32 & 61\\
       (20,400)  & 6.1 & 2.2 & 3.7 & 5.3 & & 41 & 57 & 30 & 64 & & 39 & 57 & 28 & 64\\
       (20,800)  & 5.5 & 0.8 & 4.0 & 5.6 & & 41 & 54 & 30 & 68 & & 39 & 52 & 28 & 66\\ \hline \hline
               \end{tabular}\label{t3}
           \end{table}
Next, we also consider the unequal scatter matrix case, i.e.
$\bms_1\not=\bms_2$. Now, we consider $\R_1=(0.5^{|i-j|})$ and
$\R_2=\I_p$ in this study. The other settings are all the same as
the above equal scatter matrix cases except that
$\eta=:||\D^{-1/2}(\bth_1-\bth_2)||^2/\sqrt{\tr(\R_1^2)}=0.5$. Here
we only consider two combinations of $(n_i,p)$: $(20,200),
(20,800)$. We report the simulation results in Table \ref{t4}. Our
SR test can also control the empirical sizes under this unequal
scatter matrix assumption. The other results are all similar to
Table \ref{t3}. SS is still a little conservative. Our SR still
performs better than the other tests in most cases. It shows that
our SR test can also be used in the unequal scatter matrix cases.

 \begin{table}
           \centering
           \caption{Empirical Size and  power comparison at 5\% significance with unequal scatter matrix}
           \vspace{0.1cm}
      \renewcommand{\arraystretch}{1.2}
     \tabcolsep 5pt
         \begin{tabular}{ccccccccccccccc}\hline \hline
    &  \multicolumn{4}{c}{Size}  &   & \multicolumn{4}{c}{Dense Case}&  &  \multicolumn{4}{c}{Sparse Case}     \\
          {Scenario}                 &  \multicolumn{1}{c}{CQ}  & \multicolumn{1}{c}{SS} & \multicolumn{1}{c}{PA}   & \multicolumn{1}{c}{SR}&  &  \multicolumn{1}{c}{CQ}  & \multicolumn{1}{c}{SS} & \multicolumn{1}{c}{PA}   & \multicolumn{1}{c}{SR}& &  \multicolumn{1}{c}{CQ}  & \multicolumn{1}{c}{SS} & \multicolumn{1}{c}{PA}   & \multicolumn{1}{c}{SR}  \\\hline
          \multicolumn{15}{c}{$(n_i,p)=(20,200)$} \\
(I)   & 5.8 & 2.8 & 4.2 & 6.5 && 95 & 89 & 93 & 95 && 92 & 87 & 88 & 93\\
(II)  & 6.5 & 2.5 & 3.4 & 5.5 && 58 & 91 & 93 & 95 && 52 & 87 & 88 & 93\\
(III) & 4.4 & 2.2 & 2.5 & 3.9 && 55 & 79 & 43 & 77 && 49 & 74 & 36 & 72 \\
(IV)  & 6.6 & 2.1 & 2.6 & 4.6 && 65 & 80 & 45 & 79 && 11 & 75 & 40 & 74\\
(V)   & 6.1 & 2.3 & 5.2 & 5.1 && 53 & 73 & 39 & 74 && 49 & 72 & 38 & 75\\ \hline
                 \multicolumn{15}{c}{$(n_i,p)=(20,800)$} \\
(I)   & 7.2 & 1.1 & 5.6 & 5.9 && 98 & 91 & 97 & 98 && 98 & 90 & 98 & 99\\
(II)  & 6.9 & 1.5 & 5.2 & 5.8 && 64 & 91 & 97 & 98 && 59 & 90 & 98 & 99\\
(III) & 7.3 & 1.3 & 4.6 & 4.6 && 56 & 73 & 36 & 85 && 52 & 68 & 41 & 80\\
(IV)  & 6.2 & 1.2 & 4.1 & 4.1 && 70 & 70 & 40 & 87 && 13 & 58 & 36 & 74\\
(V)   & 4.4 & 1.5 & 4.6 & 4.8 && 52 & 60 & 35 & 81 && 51 & 60 & 39 & 77\\ \hline \hline
               \end{tabular}\label{t4}
           \end{table}

Finally, to study the effect of correlation matrix on the proposed test
and to further discuss the application scope of our method, we
explore another four scenarios with different correlations and
distributions. The following moving average model is used:
\begin{align*}
X_{ijk}=||\bm
\rho_i||^{-1}(\rho_{i1}Z_{ij}+\rho_{i2}Z_{i(j+1)}+\cdots+\rho_{iT_i}
Z_{i(j+T_i-1)})+\mu_{ij}
\end{align*}
for $i=1,2$, $j=1,\cdots,n_i$ and $k=1,\cdots,p$ where $\bm
\rho_i=(\rho_{i1},\ldots,\rho_{iT_i})^T$ and $\{Z_{ijk}\}$ are
i.i.d. random variables. Consider four scenarios for the innovation
$\{Z_{ijk}\}$:
\begin{itemize}
\item[(VI)] All the $\{Z_{ijk}\}$'s are from $N(0,1)$;
\item[(VII)] the first $p/2$ components of $\{Z_{ijk}\}_{k=1}^p$ are from centralized Gamma(8,1), and the others are from $N(0,1)$.
\item[(VIII)] All the $\{Z_{ijk}\}$'s are from $t_3$;
\item[(IX)] All the $\{Z_{ijk}\}$'s are from $0.8N(0,1)+0.2N(0,9)$.
\end{itemize} The coefficients $\{\rho_{il}\}_{l=1}^{T_i}$ are
generated independently from $U(2,3)$ and are kept fixed once
generated through our simulations. The correlations among $X_{ijk}$
and $X_{ijl}$ are determined by $|k-l|$ and $T_i$. We consider the
``full dependence'' for the first sample and the ``2-dependence''
for the second sample, i.e. $T_1=p$ and $T_2=3$, to generate
different covariances of $\X_{ij}$. For simplicity, set
$\eta=:||\bth_1-\bth_2||^2/\sqrt{\tr({\bf \Lambda}_1^2)+\tr({\bf
\Lambda}_2^2)}=0.1$ where ${\bf \Lambda}_i$ is the covariance matrix
of $\X_{ij}$ and $(n_i,p)=(20,200), (20,800)$. Table \ref{t5}
reports the simulation results under these four non-elliptical
distributions. Our SR test also performs well in these cases. The
empirical sizes of SR are close to the nominal level. The power of
SR test is still a little larger than PA and SS in most cases. It
also shows the robustness of our SR test.

 \begin{table}
           \centering
           \caption{Empirical Size and  power comparison at 5\% significance with MV model}
           \vspace{0.1cm}
      \renewcommand{\arraystretch}{1.2}
     \tabcolsep 5pt
         \begin{tabular}{ccccccccccccccc}\hline \hline
    &  \multicolumn{4}{c}{Size}  &   & \multicolumn{4}{c}{Dense Case}&  &  \multicolumn{4}{c}{Sparse Case}     \\
          {Scenario}                 &  \multicolumn{1}{c}{CQ}  & \multicolumn{1}{c}{SS} & \multicolumn{1}{c}{PA}   & \multicolumn{1}{c}{SR}&  &  \multicolumn{1}{c}{CQ}  & \multicolumn{1}{c}{SS} & \multicolumn{1}{c}{PA}   & \multicolumn{1}{c}{SR}& &  \multicolumn{1}{c}{CQ}  & \multicolumn{1}{c}{SS} & \multicolumn{1}{c}{PA}   & \multicolumn{1}{c}{SR}  \\\hline
          \multicolumn{15}{c}{$(n_i,p)=(20,200)$} \\
(VI)  &6.7 &5.8 &6.4 &5.9  && 32  & 28  & 26  & 31  && 43  & 42  & 36  & 50  \\
(VII) &3.9 &4.1 &6.1 &4.8  && 23  & 32  & 31  & 33  && 38  & 72  & 69  & 76 \\
(VIII)&6.3 &3.7 &7.1 &5.1  && 41  & 34  & 35  & 41  && 42  & 44  & 34  & 53 \\
(IX)  &6.5 &5.4 &6.3 &5.7  && 29  & 27  & 26  & 33  && 34  & 36  & 28  & 47 \\ \hline
                 \multicolumn{15}{c}{$(n_i,p)=(20,800)$} \\
(VI)  &6.6 &4.8 &6.7 &5.8  && 32  & 27  & 26  & 28  && 33  & 35  & 22  & 40  \\
(VII) &4.1 &4.0 &6.3 &4.7  && 37  & 43  & 38  & 46  && 38  & 72  & 68  & 74 \\
(VIII)&5.9 &3.9 &7.8 &5.4  && 39  & 32  & 31  & 42  && 41  & 29  & 31  & 49 \\
(IX)  &6.4 &3.4 &6.2 &5.1  && 34  & 32  & 28  & 34  && 31  & 31  & 23  & 35 \\ \hline \hline
               \end{tabular}\label{t5}
           \end{table}

All the above simulation results show that our SR test is very robust and efficient test procedure in a wide range of distributions.
SR performs better than the other tests based on the direct observations when the distribution is heavy-tailed.
 In addition, when the dimension is larger than the square of sample sizes,
 our SR test is more efficient than SS test because of the conservation of SS test in this case.

\section{Discussion}
In this article, we propose a new test for high dimensional two
sample location problem based on spatial rank. Compared with the
other $L_2$-norm-based tests, our proposed test is very robust and
efficient, especially for heavy-tailed or skewed distributions. In
another direction, \cite{r3} proposed a test based on max-norm of
marginal $t$-statistics. \cite{r25} also proposed a
$L_2$-thresholding statistic. Both these two tests can detect more
sparse and stronger signals whereas the $L_2$-norm-based tests is
for denser but fainter signals. Developing a spatial-rank-based test
for sparse signals is very interest and deserves further study.

In the case of elliptical distributions, \cite{r131} propose a class
of tests based on interdirections and pseudo-Mahalanobis ranks when
the dimension is fixed, which are distribution-free,
affine-invariant, and achieve semiparametric efficiency at given
reference densities. The Hallin-Paindaveine signs and ranks have
been successful in many problems involving elliptical densities (one
and two-sample location; scatter; homogeneity of scatter;
regression; VARMA dependence; principal components, etc.). How to
construct a test base on Hallin-Paindaveine signs and ranks for high
dimensional data deserves further studies.

\end{document}


\maketitle

\section{Proofs of Theorems}
\subsection{Proof of Theorem 1}
Define $\u_{1i}=E(U(\bmv_{1i}-\bmv_{2j})|\bmv_{1i})$  and
$\u_{2j}=E(U(\bmv_{1i}-\bmv_{2j})|\bmv_{2j})$. Obviously, $\u_{1i},
\u_{2j}$ have the same distribution. And
$E(\u_{1i}\u_{1i}^T)=\tau_Fp^{-1}\I_p$ where the constant $\tau_F$
depend on the background distribution $F$. Define
\begin{align*}
\Y_{ij}=&\D^{-1/2}(\X_{ij}-\bmu_i),
\V_{2i}=-E(U(\Y_{1j}-\Y_{2i})|\Y_{2i}),\\
\V_{1i}=&E(U(\Y_{1i}-\Y_{2j})|\Y_{1i}),
P_{is}=U(\Y_{1i}-\Y_{2s})-\V_{1i}+\V_{2s},\\
\W_{ij}=&U(\Y_{1i}-\Y_{1j})-{\V}_{1i}-{\V}_{1j},\\
\A=&E(\V_{1i}\V_{1i}^T)=E(\V_{2j}\V_{2j}^T).
\end{align*}
 Let
$\D_i=\diag\{d_{i1},\cdots,d_{ip}\}$ be the diagonal matrix of
$\bms_i$ and $\hat{\D}_i=\diag\{\hat{d}_{i1},\cdots,\hat{d}_{ip}\}$.

First, we restate Lemma 4 in \citet{r26} and propose some useful
Lemmas. The proof of these Lemmas are given in Appendix B.
\begin{lemma} \label{le1}
Suppose $\u$ are independent identically distributed uniform on the
unit $p$ sphere. For any $p\times p$ symmetric matrix $\M$, we have
\begin{align*}
E(\u^T\M\u)^2=&\{\tr^2(\M)+2\tr(\M^2)\}/(p^2+2p),\\
E(\u^T\M\u)^4=&\{3\tr^2(\M^2)+6\tr(\M^4)\}/\{p(p+2)(p+4)(p+6)\}.
\end{align*}
\end{lemma}

\begin{lemma}\label{le2}
Under Condition (C4), we have $\max_{1\le j \le
p}(\hat{d}_{ij}-d_{ij})=O_p(n_i^{-1/2}(\log p)^{1/2})$.
\end{lemma}

\begin{lemma}\label{le3}
$\tau_F\to 0.5$ as $p\to \infty$.
\end{lemma}

\begin{lemma}\label{le4}
Suppose the conditions given in Theorem 1 all hold, we have
$T_n=Z_n+o_p(\sigma_n)$, where
\begin{align*}
Z_n=&\frac{1}{n_1(n_1-1)}\underset{i\not=j}{\sum^{n_1}\sum^{n_1}}\V_{1i}^T\V_{1j}
+\frac{1}{n_2(n_2-1)}\underset{i\not=j}{\sum^{n_2}\sum^{n_2}}\V_{2i}^T\V_{2j}\\
&+\frac{2}{n_1n_2}\sum_{i=1}^{n_1}\sum_{j=1}^{n_2}\V_{1i}\V_{2j}.
\end{align*}
\end{lemma}

\begin{lemma}\label{le5}
Suppose the conditions given in Theorem 1 all hold. Then,
${Z_n}/{\sigma_n}\cd N(0,1)$.
\end{lemma}

\vspace{0.5cm} \noindent{\it Proof of Theorem 1}: According to Lemma
\ref{le4} and \ref{le5}, we can easily obtain the result.
\hfill$\Box$

\subsection{Proof of Proposition 1}
Taking the same procedure as Theorem 1, we can obtain
\begin{align*}
\widehat{\tr(\R^2)}_1=&\frac{2p^2}{P_{n_1}^4}\sum^*U(\Y_{1i_1}-\Y_{1i_2})^T U(\Y_{1i_3}-\Y_{1i_4})U(\Y_{1i_3}-\Y_{1i_2})^TU(\Y_{1i_1}-\Y_{1i_4})+o_p(\tr(\R^2))\\
=&\frac{2p^2}{P_{n_1}^4}\sum^*(\V_{1i_1}-\V_{1i_2})^T(\V_{1i_3}-\V_{1i_4})(\V_{1i_3}-\V_{1i_2})^T(\V_{1i_1}-\V_{1i_4})+o_p(\tr(\R^2))\\
=&\frac{2p^2}{P_{n_1}^2}\sum^*(\V_{1i_1}^T\V_{1i_2})^2-\frac{4p^2}{P_{n_1}^3}\sum^*\V_{1i_1}^T\V_{1i_2}\V_{1i_2}^T\V_{1i_3}\\
&+\frac{2p^2}{P_{n_1}^4}\sum^*\V_{1i_1}^T\V_{1i_2}\V_{1i_3}^T\V_{1i_4}+o_p(\tr(\R^2))\\
\doteq & J_1+J_2+J_3+o_p(\tr(\R^2)).
\end{align*}
Obviously, $E(J_1)=4p^2\tr(\A^2)$. And
\begin{align*}
\tr(\A^2)=E\{(\V_{1i_1}^T\V_{1i_2})^2\}=&E\left[\frac{(\u_{1i_1}^T\R\u_{1i_2})^2}{\{1+\u_{1i}^T(\R-\I_p)\u_{1j}\}^2}\right]\\
=&E\left\{(\u_{1i_1}^T\R\u_{1i_2})^2\right\}\{1+o_p(1)\}\\
=&\tau_F^2p^{-2}\tr(\R^2)\{1+o(1)\}=4^{-1}p^{-2}\tr(\R^2)\{1+o(1)\},
\end{align*}
because
$E[\left\{\u_{1i}^T(\R-\I_p)\u_{1j}\right\}^2]=\tau_F^2p^{-2}(\tr(\R^2)-\I_p)=o(1)$
by Condition (C3). So $E(J_1)=\tr(\R^2)\{1+o(1)\}$. Taking the same
procedure as above, we have
\begin{align*}
\var&\left(\frac{2p^2}{n_1(n_1-1)}\sum_{i\not=j}(\V_{1i}^T\V_{1j})^2\right)\\
=&O(n_1^{-2}p^4)E((\V_{1i}^T\V_{1j})^4)+O(n_1^{-1}p^4)[E\{(\V_{1i}^T\V_{1j})^2\}]^2\\
=&O(n_1^{-2}\tr(\R^4)+n_1^{-1}\tr^2(\R^2))=o(\tr^2(\R^2)).
\end{align*}
by Lemma \ref{le1} and Condition (C2). Similarly, we can show that
\begin{align*}
E(J_2^2)=&O(n_1^{-3}p^4)E((\V_{1i}^T\A\V_{1i})^2)+O(n_1^{-2}p^4)\tr(\A^4)=o(\tr^2(\R^2)),\\
E(J_3^2)=&O(n_1^{-4}p^4\tr(\A^4))=o(\tr^2(\R^2)).
\end{align*}
Thus, $\widehat{\tr(\R^2)}_1=\tr(\R^2)(1+o_p(1))$. We can also show
the ratio-consistency of the other estimators.\hfill$\Box$

\subsection{Proof of Theorem 2}
Define $\U_{is}=U(\Y_{1i}-\Y_{2s})$ and
$r_{is}=||\Y_{1i}-\Y_{2s}||$. Firstly, taking the same procedure as
Lemma \ref{le4}, we can show that
\begin{align*}
T_n=&\frac{1}{n_1(n_1-1)}\frac{1}{n_2(n_2-1)}
\underset{i\not=j}{\sum^{n_1}\sum^{n_1}}
 \underset{s\not=t}{\sum^{n_2}\sum^{n_2}}U(\D^{-1/2}(\X_{1i}-\X_{2s}))^TU(\D^{-1/2}(\X_{1j}-\X_{2t}))+o_p(\sigma_n)\\
=&\frac{1}{n_1(n_1-1)}\frac{1}{n_2(n_2-1)}
\underset{i\not=j}{\sum^{n_1}\sum^{n_1}}
 \underset{s\not=t}{\sum^{n_2}\sum^{n_2}}U(\Y_{1i}-\Y_{2s}+\D^{-1/2}(\bmu_1-\bmu_2))^T\\
 &\times U(\Y_{1j}-\Y_{2t}+\D^{-1/2}(\bmu_1-\bmu_2))+o_p(\sigma_n)\\
 =&\frac{1}{n_1(n_1-1)}\frac{1}{n_2(n_2-1)}  \underset{i\not=j}{\sum^{n_1}\sum^{n_1}}
 \underset{s\not=t}{\sum^{n_2}\sum^{n_2}}U(\Y_{1i}-\Y_{2s})^TU(\Y_{1j}-\Y_{2t})\\
 &+\frac{2}{n_1(n_1-1)}\frac{1}{n_2(n_2-1)}  \underset{i\not=j}{\sum^{n_1}\sum^{n_1}}
 \underset{s\not=t}{\sum^{n_2}\sum^{n_2}}r_{is}^{-1}\U_{jt}^T\left[\I_p-\U_{is}\U_{is}^T\right]\D^{-1/2}(\bmu_1-\bmu_2)\\
 &+\frac{1}{n_1(n_1-1)}\frac{1}{n_2(n_2-1)}  \underset{i\not=j}{\sum^{n_1}\sum^{n_1}}
 \underset{s\not=t}{\sum^{n_2}\sum^{n_2}}r_{is}^{-1}r_{jt}^{-1}(\bmu_1-\bmu_2)^T\D^{-1/2}\left[\I_p-\U_{jt}\U_{jt}^T\right]\\
 &\times
 \left[\I_p-\U_{is}\U_{is}^T\right]\D^{-1/2}(\bmu_1-\bmu_2)+o_p(\sigma_n).
\end{align*}
According to the same arguments as Theorem 1, we have
\begin{align*}
\frac{2}{n_1(n_1-1)}&\frac{1}{n_2(n_2-1)}
\underset{i\not=j}{\sum^{n_1}\sum^{n_1}}
 \underset{s\not=t}{\sum^{n_2}\sum^{n_2}}r_{is}^{-1}\U_{jt}^T\left[\I_p-\U_{is}\U_{is}^T\right]\D^{-1/2}(\bmu_1-\bmu_2)\\
 =&\frac{2c_0}{n_1}\sum_{i=1}^{n_1}\V_{1i}^T\D^{-1/2}(\bmu_1-\bmu_2)-\frac{2c_0}{n_2}\sum_{i=1}^{n_2}\V_{2i}^T\D^{-1/2}(\bmu_1-\bmu_2)+o_p(\sigma_n),\\
 \frac{2}{n_1(n_1-1)}&\frac{1}{n_2(n_2-1)}  \underset{i\not=j}{\sum^{n_1}\sum^{n_1}}
 \underset{s\not=t}{\sum^{n_2}\sum^{n_2}}r_{is}^{-1}r_{jt}^{-1}(\bmu_1-\bmu_2)^T\D^{-1/2}\left[\I_p-\U_{jt}\U_{jt}^T\right]\\
 &\times \left[\I_p-\U_{is}\U_{is}^T\right]\D^{-1/2}(\bmu_1-\bmu_2)\\
 =&c_0^2(\bmu_1-\bmu_2)^T\D^{-1}(\bmu_1-\bmu_2)+o_p(\sigma_n).
\end{align*}
where $c_0=E(r_{is}^{-1})$. Thus,
\begin{align*}
T_n=&Z_n+\frac{2c_0}{n_1}\sum_{i=1}^{n_1}\V_{1i}^T\D^{-1/2}(\bmu_1-\bmu_2)-\frac{2c_0}{n_2}\sum_{i=1}^{n_2}\V_{2i}^T\D^{-1/2}(\bmu_1-\bmu_2)\\
&+c_0^2(\bmu_1-\bmu_2)^T\D^{-1}(\bmu_1-\bmu_2)+o_p(\sigma_n).
\end{align*}
And
\begin{align*}
E\left(\frac{2c_0}{n_1}\sum_{i=1}^{n_1}\V_{1i}^T\D^{-1/2}(\bmu_1-\bmu_2)\right)^2=&O(n^{-1}p^{-1}c_0^2(\bmu_1-\bmu_2)^T\D^{-1/2}\R\D^{-1/2}(\bmu_1-\bmu_2)),\\
E\left(\frac{2c_0}{n_2}\sum_{i=1}^{n_2}\V_{2i}^T\D^{-1/2}(\bmu_1-\bmu_2)\right)^2=&O(n^{-1}p^{-1}c_0^2(\bmu_1-\bmu_2)^T\D^{-1/2}\R\D^{-1/2}(\bmu_1-\bmu_2)).
\end{align*}
Under Condition (C4), the second and third parts of $T_n$ are all
$o_p(\sigma_n)$. Then, by Theorem 1, we have
\begin{align*}
\frac{T_n-c_0^2(\bmu_1-\bmu_2)^T\D^{-1}(\bmu_1-\bmu_2)}{\sigma_n}\cd
N(0,1).
\end{align*}
Here we complete the proof. \hfill$\Box$

\section{Proof of Lemmas}
\subsection{Proof of Lemma \ref{le2}}
\proof
\begin{align*}
R(\D^{-1/2}\X_{1i})=&\frac{1}{n_1}\sum_{j=1}^{n_1}U(\Y_{1i}-\Y_{1j})=\frac{1}{n_1}\sum_{j=1}^{n_1}(\V_{1i}-\V_{1j}+\W_{ij})\\
=&\V_{1i}-\frac{1}{n_1}\sum_{j=1}^{n_1}\V_{2j}+\frac{1}{n_1}\sum_{j=1}^{n_1}\W_{ij}.
\end{align*}
Obviously,
$\frac{1}{n_1}\sum_{j=1}^{n_1}\V_{2j}=O_p(\sqrt{n^{-1}\tr(\A^2)})$
and
$\frac{1}{n_1}\sum_{j=1}^{n_1}\W_{ij}=O_p(\sqrt{n^{-1}\tr(\A^2)})$.
Thus,
\begin{align*}
E(R(&\D^{-1/2}\X_{1i})R(\D^{-1/2}\X_{1i})^T)\\
=&E\left(\V_{1i}\V_{1i}^T\right)(1+o(1))\\
=&E(E(U(\R^{1/2}(\bmv_{1i}-\bmv_{1j}))U(\R^{1/2}(\bmv_{1i}-\bmv_{1j}))^T|\R^{1/2}\bmv_{1i}))(1+o(1))\\
=&E\left(\frac{\R^{1/2}\u_{1i}\u_{1i}^T\R^{1/2}}{1+\u_{1i}^T(\R-\I_p)\u_{1i}}\right)(1+o(1)).
\end{align*}
Thus, by the Cauchy inequality and Tyler expansion,
\begin{align*}
E&\left(\diag\left\{E\left(R(\D^{-1/2}\X_{1i})R(\D^{-1/2}\X_{1i})^T\right)\right\}-\tau_F p^{-1}\I_p\right)\\
&\le  C_4 \{E(\u_{ij}^T (\R-\I_p)\u_{ij})^2 E(\diag\{\R^{1/2}\u_{1i}\u_{1i}^T\R^{1/2}\}-\tau_Fp^{-1}\I_p)^2\}^{1/2}\\
&=O\left(p^{-1}\sqrt{\tr (\R^2)-p}\right)=o(n^{-1/2}).
\end{align*}
by Condition (C3). The above equation define the functional equation
for each component of $\bm d_i=(d_{i1},\cdots,d_{ip})$,
\begin{align}\label{tfe}
T_{ij}(F,d_{ij})=o_p(n^{-1/2}),
\end{align}where $F_i$ is the distribution function of $\X_{ij}$, $i=1,2$. Similar to \cite{r15}, the linearisation of this equation produces
\begin{align*}
\sqrt{n_i}(\hat{d}_{ij}-d_{ij})=-{\bf
H}_{ij}^{-1}\sqrt{n_i}(T_{ij}(F_{ni},d_{ij})-T_{ij}(F_i,d_{ij}))+o_p(1),
\end{align*}
where $F_{ni}$ is the empirical distribution function of $\X_{ij},
j=1,\cdots,n_i$, ${\bf H}_{ij}$ is the corresponding Hessian matrix
of the functional defined in (\ref{tfe}), and
\begin{align*}
T_i(F_{ni},\bm
d_i)=\left(vec\left(\diag\left(n_i^{-1}\sum_{j=1}^{n_i}
R(\D^{-1/2}\X_{1i})R(\D^{-1/2}\X_{1i})^T-\tau_Fp^{-1}\I_p\right)\right)\right),
\end{align*}
 where $T_i(F_{ni},\bm d_i)=(T_{i1}(F_{ni},d_{i1}),\cdots,T_{ip}(F_{ni},d_{ip}))$ and $vec(\B)$ means
the vector of the diagonal matrix of $\B$. For each variance
estimator $\hat{d}_{ij}$, we have
\begin{align*}
\sqrt{n_i}(\hat{d}_{ij}-d_{ij}) \cp N(0,\zeta_{ij}^2),
\end{align*}
where $\zeta_{ij}^2$ is the corresponding asymptotic variance.
Define $\zeta_{max}=\max_{1\le i \le 2,1\le j \le p}\zeta_{ij}$.
\begin{align*}
P&\left(\max_{1\le j \le p}(\hat{d}_{ij}-d_{ij})> \sqrt{2}\zeta_{max}n_i^{-1/2}(\log p)^{1/2}\right) \\
&\le \sum_{j=1}^p P\left(\sqrt{n_i}(\hat{d}_{ij}-d_{ij})> \sqrt{2}\zeta_{max}(\log p)^{1/2}\right)\\
&=\sum_{i=1}^p \left(1-\Phi(\sqrt{2}\zeta_{max}\sigma_{ij}^{-1}(\log p)^{1/2})\right)\le p\left(1-\Phi((2\log p)^{1/2} )\right)\\
&\le \frac{p}{\sqrt{4\pi\log p}}e^{-\log p}=(4\pi)^{-1/2}(\log
p)^{-1/2} \to 0.
\end{align*}
Finally, $\max_{1\le j \le
p}(\hat{d}_{ij}-d_{ij})=O_p(n_i^{-1/2}(\log p)^{1/2})$. \hfill$\Box$

\subsection{Proof of Lemma \ref{le3}}
\proof
\begin{align*}
E(\bmv_i^T\bmv_i)=&E((\bmv_i-\bmv_j)^T(\bmv_i-\bmv_k))\\
=&E(E((\bmv_i-\bmv_j)^T(\bmv_i-\bmv_k)\big|\bmv_i))\\
=&E(E(||\bmv_i-\bmv_j||||\bmv_i-\bmv_k||U(\bmv_i-\bmv_j)^TU(\bmv_i-\bmv_k)\big|\bmv_i))\\
=&E((E(||\bmv_i-\bmv_j||\big|\bmv_i))^2)E(E(U(\bmv_i-\bmv_j)^TU(\bmv_i-\bmv_k)\big|\bmv_i))\\
=&E((E(||\bmv_i-\bmv_j||\big|\bmv_i))^2)E(u_i^Tu_i)=\tau_FE((E(||\bmv_i-\bmv_j||\big|\bmv_i))^2).
\end{align*}
In addition, $E(||\bmv_i||^2)=0.5E(||\bmv_i-\bmv_j||^2)$. Thus, we
only need to show that
\begin{align*}
\frac{E((E(||\bmv_i-\bmv_j||\big|\bmv_i))^2)}{E(||\bmv_i-\bmv_j||^2)}\to
1.
\end{align*}
Because $\bmv_i$ has the elliptical distribution, $\bmv_i-\bmv_j$
also has the elliptical distribution. Define the density function of
$||\bmv_i-\bmv_j||$ is $f(t)=c_pt^{p-1}g(t)$ where
$c_p=\frac{2\pi^{p/2}}{\Gamma(p/2)}$. Thus,
\begin{align*}
\frac{E((E(||\bmv_i-\bmv_j||\big|\bmv_i))^2)}{E(||\bmv_i-\bmv_j||^2)}=&\frac{\left(\int c_p t^p g(t) dt\right)^2}{\int c_p t^{p+1} g(t) dt}\\
=&\frac{c_{p+1}^2}{c_pc_{p+2}}=\frac{\Gamma^2((p+1)/2)}{\Gamma(p/2)\Gamma((p+2)/2)}.
\end{align*}
By the Stirling's formula,
\[
\lim_{x \rightarrow \infty}\frac{\Gamma(x+1)}{(x/e)^{x}(2\pi
x)^{1/2}}=1,\] as $p\to \infty$, we have
\begin{align*}
\frac{c_{p+1}^2}{c_pc_{p+2}} \to
\frac{(p-1)^{p-1}}{p^{p/2}(p-2)^{(p-2)/2}}=(1-p^{-1})^{p/2}(1+(p-2)^{-1})^{(p-2)/2}
\to 1.
\end{align*}
Here we complete the proof. \hfill$\Box$

\subsection{Proof of Lemma \ref{le4}}
\proof By the Taylor's expansion, we have
\begin{align*}
U(\hat{\D}_{(i,j,s,t)}^{-1/2}&(\X_{1i}-\X_{2s}))^TU(\hat{\D}_{(i,j,s,t)}^{-1/2}(\X_{1j}-\X_{2t}))\\
=&U(\D^{-1/2}(\X_{1i}-\X_{2s}))^TU(\D^{-1/2}(\X_{1j}-\X_{2t}))\\
&+U(\D^{-1/2}(\X_{1j}-\X_{2t}))^T(\I_p-U(\D^{-1/2}(\X_{1i}-\X_{2s}))U(\D^{-1/2}(\X_{1i}-\X_{2s}))^T)\\
&\times(\hat{\D}_{(i,j,s,t)}^{-1/2}\D^{1/2}-\I_p)U(\D^{-1/2}(\X_{1i}-\X_{2s}))\\
&+U(\D^{-1/2}(\X_{1i}-\X_{2s}))^T(\I_p-U(\D^{-1/2}(\X_{1j}-\X_{2t}))U(\D^{-1/2}(\X_{1j}-\X_{2t}))^T)\\
&\times(\hat{\D}_{(i,j,s,t)}^{-1/2}\D^{1/2}-\I_p)U(\D^{-1/2}(\X_{1j}-\X_{2t}))\\
&+o_p(\sigma_n).
\end{align*}
Next, we only show that
\begin{align*}
\frac{1}{n_1(n_1-1)}\frac{1}{n_2(n_2-1)} &
\underset{i\not=j}{\sum^{n_1}\sum^{n_1}}
 \underset{s\not=t}{\sum^{n_2}\sum^{n_2}}U(\D^{-1/2}(\X_{1j}-\X_{2t}))^T(\hat{\D}_{(i,j,s,t)}^{-1/2}\D^{1/2}-\I_p)\\
 &\times U(\D^{-1/2}(\X_{1i}-\X_{2s}))=o_p(\sigma_n).
\end{align*}
Similarly, by
$U(\D^{-1/2}(\X_{1i}-\X_{2s}))=\V_{1i}-\V_{2s}+P_{is}$, we have
\begin{align*}
\frac{1}{n_1(n_1-1)}&\frac{1}{n_2(n_2-1)}
\underset{i\not=j}{\sum^{n_1}\sum^{n_1}}
 \underset{s\not=t}{\sum^{n_2}\sum^{n_2}}U(\D^{-1/2}(\X_{1j}-\X_{2t}))^T(\hat{\D}_{(i,j,s,t)}^{-1/2}\D^{1/2}-\I_p)\\
 &\times U(\D^{-1/2}(\X_{1i}-\X_{2s}))\\
 &=\frac{1}{n_1(n_1-1)}\underset{i\not=j}{\sum^{n_1}\sum^{n_1}}\V_{1i}^T(\hat{\D}_{(i,j,s,t)}^{-1/2}\D^{1/2}-\I_p)\V_{1j}\\
&+\frac{1}{n_2(n_2-1)}\underset{i\not=j}{\sum^{n_2}\sum^{n_2}}\V_{2i}^T(\hat{\D}_{(i,j,s,t)}^{-1/2}\D^{1/2}-\I_p)\V_{2j}\\
&+\frac{2}{n_1n_2}\sum_{i=1}^{n_1}\sum_{j=1}^{n_2}\V_{1i}^T(\hat{\D}_{(i,j,s,t)}^{-1/2}\D^{1/2}-\I_p)\V_{2j}\\
&+\frac{2}{n_1n_2(n_1-1)}\underset{i\not=j}{\sum^{n_1}}\sum_{s=1}^{n_2}P_{is}^T(\hat{\D}_{(i,j,s,t)}^{-1/2}\D^{1/2}-\I_p)\V_{1j}\\
&+\frac{2}{n_2n_1(n_2-1)}\underset{s\not=t}{\sum^{n_2}}\sum_{i=1}^{n_1}P_{is}^T(\hat{\D}_{(i,j,s,t)}^{-1/2}\D^{1/2}-\I_p)\V_{2t}\\
&+\frac{1}{n_1(n_1-1)}\frac{1}{n_2(n_2-1)}
\underset{i\not=j}{\sum^{n_1}\sum^{n_1}}
 \underset{s\not=t}{\sum^{n_2}\sum^{n_2}}P_{is}^T(\hat{\D}_{(i,j,s,t)}^{-1/2}\D^{1/2}-\I_p)P_{jt}\\
&\doteq Q_1+Q_2+Q_3+Q_4+Q_5+Q_6.
\end{align*}
Here we only proof $E(Q_1^2)=o(\sigma_n^2)$. The other parts $Q_2,
Q_3$ are similar to $Q_1$. And the last three past are similar to
the following proof of $T_1$. Taking the same arguments as Lemma
\ref{le2},
\begin{align*}
E(Q_1^2)=&\frac{1}{n_1(n_1-1)}E\left(\V_{1i}^T(\hat{\D}_{(i,j,s,t)}^{-1/2}\D^{1/2}-\I_p)\V_{1j}\right)^2\\
=&\frac{1}{n_1(n_1-1)}E\left(\u_{1i}^T\R^{1/2}(\hat{\D}_{(i,j,s,t)}^{-1/2}\D^{1/2}-\I_p)\R^{1/2}\u_{1j}\right)^2(1+o(1))\\
=&\frac{1}{n_1(n_1-1)}\tr(\A^2)(\log(p)/n)(1+o(1))=o(\sigma_n^2).
\end{align*}
Thus, under $H_0$, we have
\begin{align*}
T_n=&\frac{1}{n_1(n_1-1)}\frac{1}{n_2(n_2-1)}
\underset{i\not=j}{\sum^{n_1}\sum^{n_1}}
 \underset{s\not=t}{\sum^{n_2}\sum^{n_2}}(\V_{1i}+\V_{2s}+P_{is})^T(\V_{1j}+\V_{2t}+P_{jt})+o_p(\sigma_n)\\
 =&\frac{1}{n_1(n_1-1)}\underset{i\not=j}{\sum^{n_1}\sum^{n_1}}\V_{1i}^T\V_{1j}
+\frac{1}{n_2(n_2-1)}\underset{i\not=j}{\sum^{n_2}\sum^{n_2}}\V_{2i}^T\V_{2j}\\
&+\frac{2}{n_1n_2}\sum_{i=1}^{n_1}\sum_{j=1}^{n_2}\V_{1i}^T\V_{2j}+\frac{2}{n_1n_2(n_1-1)}\underset{i\not=j}{\sum^{n_1}}\sum_{s=1}^{n_2}P_{is}^T\V_{1j}\\
&+\frac{2}{n_2n_1(n_2-1)}\underset{s\not=t}{\sum^{n_2}}\sum_{i=1}^{n_1}P_{is}^T\V_{2t}+\frac{1}{n_1(n_1-1)}\frac{1}{n_2(n_2-1)}
\underset{i\not=j}{\sum^{n_1}\sum^{n_1}}
 \underset{s\not=t}{\sum^{n_2}\sum^{n_2}}P_{is}^TP_{jt}+o_p(\sigma_n)\\
\doteq &Z_n+T_1+T_2+T_3+o_p(\sigma_n).
\end{align*}
Next, we only show that $T_1=o_p(\sigma_n)$. $T_2$ and $T_3$ are
similar to $T_1$.
\begin{align*}
E(T_1^2)=&E\left(\frac{2}{n_1n_2(n_1-1)}\underset{i\not=j}{\sum^{n_1}}\sum_{s=1}^{n_2}P_{is}^T\V_{1j}\right)^2\\
=&\frac{4n_1(n_1-1)n_2(n_2-1)}{n_1^2n_2^2(n_1-1)^2}E(P_{is}^T\V_{1j}P_{it}^T\V_{1j})\\
&+\frac{4n_1(n_1-1)n_2(n_2-1)}{n_1^2n_2^2(n_1-1)^2}E(P_{is}^T\V_{1j}P_{jt}^T\V_{1i})\\
&+\frac{4n_1(n_1-1)(n_1-2)n_2}{n_1^2n_2^2(n_1-1)^2}E(P_{is}^T\V_{1j}P_{ks}^T\V_{1j})\\
&+\frac{4n_1(n_1-1)(n_1-2)n_2}{n_1^2n_2^2(n_1-1)^2}E(P_{is}^T\V_{1j}P_{ks}^T\V_{1i})\\
&+\frac{4}{n_1n_2(n_1-1)}E(P_{is}^T\V_{1j}\V_{1j}^TP_{is})+\frac{4}{n_1n_2(n_1-1)}E(P_{is}^T\V_{1j}P_{js}^T\V_{1i})\\
=&\frac{4}{n_1n_2(n_1-1)}E(P_{is}^T\V_{1j}\V_{1j}^TP_{is})+\frac{4}{n_1n_2(n_1-1)}E(P_{is}^T\V_{1j}P_{js}^T\V_{1i}),
\end{align*}
because
\begin{align*}
E(P_{is}^T&\V_{1j}P_{it}^T\V_{1j})=E(P_{is}^T\A P_{it})\\
=&E((U(\Y_{1i}-\Y_{2s})-\V_{1i}-\V_{2s})^T\A (U(\Y_{1i}-\Y_{2t})-\V_{1i}-\V_{2t}))\\
=&E(U(\Y_{1i}-\Y_{2s})^T\A U(\Y_{1i}-\Y_{2t}))-E(U(\Y_{1i}-\Y_{2s})^T\A \V_{1i})\\
&-E(U(\Y_{1i}-\Y_{2t})^T\A \V_{1i})+E(\V_{1i}^T\A \V_{1i})\\
=&E(E(U(\Y_{1i}-\Y_{2s})^T\A U(\Y_{1i}-\Y_{2t})|\Y_{1i}))-E(E(U(\Y_{1i}-\Y_{2s})^T\A \V_{1i})|\Y_{1i})\\
&-E(E(U(\Y_{1i}-\Y_{2t})^T\A \V_{1i})|\Y_{1i})+\tr(\A ^2)\\
=&E(\V_{1i}^T\A \V_{1i})-E(\V_{1i}^T\A \V_{1i})-E(\V_{1i}^T\A \V_{1i})+\tr(\A ^2)\\
=&0,\\
E(P_{is}^T&\V_{1j}P_{ks}^T\V_{1j})=E(P_{is}^T\A P_{ks})=0,\\
E(P_{is}^T&\V_{1j}P_{jt}^T\V_{1i})=\tr(E(\V_{1i}P_{is}^T))^2,\\
E(P(&\Y_{1i})P_{is}^T)=E(\V_{1i}(U(\Y_{1i}-\Y_{2s})-\V_{1i}-\V_{2s}))\\
&=E(\V_{1i}U(\Y_{1i}-\Y_{2s})^T)-E(\V_{1i}\V_{1i}^T)\\
&=E(E(\V_{1i}U(\Y_{1i}-\Y_{2s})^T|\Y_{1i}))-E(\V_{1i}\V_{1i}^T)\\
&=E(\V_{1i}\V_{1i}^T)-E(\V_{1i}\V_{1i}^T)=0.
\end{align*}
{ Next, we will show that
$E(P_{is}^T\V_{1j}\V_{1j}^TP_{is})=E(P_{is}^T\A
P_{is})=O(\tr(\A^2))$.} In fact, we only need to show that
$E(U(\Y_{1i}-\Y_{2j})^T\A U(\Y_{1i}-\Y_{2j}))=O(\tr(\A^2))$.
\begin{align*}
U(\Y_{1i}-&\Y_{2j})^T\A U(\Y_{1i}-\Y_{2j})\\
=&\frac{\Y_{1i}-\Y_{2j}}{||\Y_{1i}-\Y_{2j}||}\A U(\Y_{1i}-\Y_{2j})\\
=&\frac{\Y_{1i}-\Y_0+\Y_0-\Y_{2j}}{||\Y_{1i}-\Y_{2j}||}\A U(\Y_{1i}-\Y_{2j})\\
=&\frac{\Y_{1i}-\Y_0}{||\Y_{1i}-\Y_{2j}||}\A U(\Y_{1i}-\Y_{2j})+\frac{\Y_0-\Y_{2j}}{||\Y_{1i}-\Y_{2j}||}\A U(\Y_{1i}-\Y_{2j})\\
=&U(\Y_{1i}-\Y_0)^T\A U(\Y_{1i}-\Y_{2j}) \frac{||\Y_{1i}-\Y_0||}{||\Y_{1i}-\Y_{2j}||}\\
&+U(\Y_0-\Y_{2j})^T\A U(\Y_{1i}-\Y_{2j})
\frac{||\Y_0-\Y_{2j}||}{||\Y_{1i}-\Y_{2j}||}.
\end{align*}
Additionally,
\begin{align*}
&E\left(U(\Y_{1i}-\Y_0)\A U(\Y_{1i}-\Y_{2j}) \frac{||\Y_{1i}-\Y_0||}{||\Y_{1i}-\Y_{2j}||}\right)\\
&=E\left(U(\bmv_{1i}-\bmv_0)^T\R^{1/2}\A \R^{1/2} U(\bmv_{1i}-\bmv_{2j})\frac{||\bmv_{1i}-\bmv_0||}{||\bmv_{1i}-\bmv_{2j}||}\right)(1+o(1))\\
&=E\left(U(\bmv_{1i}-\bmv_0)^T\R^{1/2}\A \R^{1/2} U(\bmv_{1i}-\bmv_{2j})\frac{||\bmv_{1i}-\bmv_0||}{||\bmv_{1i}-\bmv_{2j}||}\bigg|\bmv_{1i}\right)(1+o(1))\\
&=E\left(\u_{1i}^T\R^{1/2}\A\R^{1/2}\u_{1i}\right) E\left(\frac{||\bmv_{1i}-\bmv_0||}{||\bmv_{1i}-\bmv_{2j}||}\right)(1+o(1))\\
&=O(\tr^2(\A^2)).
\end{align*}
Similarly, we can show another part is also $O(\tr(\A^2))$. Thus,
$E(U(\Y_{1i}-\Y_{2j})^T\A U(\Y_{1i}-\Y_{2j}))=O(\tr(\A^2))$. Then,
we obtain that $T_n=Z_n+o_p(\sigma_n)$. \hfill$\Box$

\subsection{Proof of Lemma \ref{le5}}
\proof Let $\U_i=\V_{1i}$ for $i=1,\ldots,n_1$ and
$\U_{j+n1}=\V_{2j}$ for $j=1,\ldots,n_2$ and for $i\not=j$,
\begin{align*}
\phi_{ij}=\left\{\hspace{-0.1cm}\begin{array}{ll} n_1^{-1}(n_1-1)^{-1}\U_i^{T}\U_j, &  i,j \in \{1,2,\ldots,n_1\},\\[0.1cm]
-n_1^{-1}n_2^{-1}\U_i^{T}\U_j, & i\in\{1,2,\ldots,n_1\},
j\in\{n_1+1,\ldots,n\},\\[0.1cm]
n_2^{-1}(n_2-1)^{-1}\U_i^{T}\U_j, &  i,j \in
\{n_1+1,n_1+2,\ldots,n\}.
\end{array}\right.
\end{align*}
Define $Z_{nj}=\sum_{i=1}^{j-1}\phi_{ij}$ for $j=2,3,\ldots,n$,
$S_{nm}=\sum_{j=1}^m Z_{nj}$ and $\mathcal
{F}_{nm}=\sigma\{\U_1,\U_2,\ldots,\U_m\}$ which is the
$\sigma$-algebra generated by $\{\U_1,\U_2,\ldots,\U_m\}$. Now
\[Z_n=2\sum_{j=2}^{n}Z_{nj}.\]
We can verify that for each $n$,
$\{S_{nm},\mathcal{F}_{nm}\}_{m=1}^n$ is the sequence of zero mean
and a square integrable martingale. In order to prove the normality
of $Z_n$, according to Hall and Heyde (1980), it suffices to show
the following two results:

\begin{align}
&\frac{\sum_{j=2}^{n}E[Z_{nj}^2|\mathcal{F}_{n,j-1}]}{\sigma_n^2}\cp
\frac{1}{4}\label{lee3},\\
&\sigma_n^{-2}\sum_{j=2}^{n}E[Z_{nj}^2I(|Z_{nj}|>
\epsilon\sigma_n|)|\mathcal{F}_{n,j-1}]\cp 0.\label{lee4}
\end{align}
First, we proof result (\ref{lee3}). Note that
\begin{align*}
E[Z_{nj}^2|\mathcal{F}_{n,j-1}]&=\frac{1}{\tilde{n}_j^2(\tilde{n}_j-1)^2}E\left\{\left(\sum_{i=1}^{j-1}\U_i^{T} \U_j\right)^2|\mathcal{F}_{n,j-1}\right\}\\
&=\frac{1}{\tilde{n}_j^2(\tilde{n}_j-1)^2}E\left\{\sum_{i_1,i_2=1}^{j-1}\U_{i_1}^{T}\U_j\U_j^{T}\U_{i_2}|\mathcal{F}_{n,j-1}\right\}\\
&=\frac{1}{\tilde{n}_j^2(\tilde{n}_j-1)^2}\sum_{i_1,i_2=1}^{j-1}\U_{i_1}^{T}E\left(\U_j\U_j^{T}|\mathcal{F}_{n,j-1}\right)\U_{i_2}\\
&=\frac{1}{\tilde{n}_j^2(\tilde{n}_j-1)^2}\sum_{i_1,i_2=1}^{j-1}\U_{i_1}^{T}\A\U_{i_2},
\end{align*}
where $\tilde{n}_j=n_1$, for $j \in [1,n_1]$ and $\tilde{n}_j=n_2$,
for $j \in [n_1,n]$. Define
$\eta_n=\sum_{j=2}^{n}E[Z_{nj}^2|\mathcal{F}_{n,j-1}]$. By some
tedious algebra, we can obtain that
$E(\eta_n)=\frac{1}{4}\sigma_n^2(1+o(1))$.

Now write $E(\eta_n^2)$ as
\begin{align*}
E(\eta_n^2)=&E\left\{\sum_{j=2}^{n}\frac{1}{\tilde{n}_j^2(\tilde{n}_j-1)^2}\sum_{i_1,i_2=1}^{j-1}\U_{i_1}^{T}\A\U_{i_2}\right\}^2\\
=&2E\left\{\sum_{2\le
j_1<j_2}^{n}\frac{1}{\tilde{n}_{j_1}^2(\tilde{n}_{j_1}-1)^2}\frac{1}{\tilde{n}_{j_2}^2(\tilde{n}_{j_2}-1)^2}\sum_{i_1,i_2=1}^{j_1-1}\sum_{i_3,i_4=1}^{j_2-1}
\U_{i_1}^{T}\A\U_{i_2}\U_{i_3}^{T}\A\U_{i_4}\right\}\\
&+E\left\{\sum_{j=2}^{n}\frac{1}{\tilde{n}_{j}^4(\tilde{n}_{j}-1)^4}\sum_{i_1,i_2=1}^{j-1}\sum_{i_3,i_4=1}^{j-1}
\U_{i_1}^{T}\A\U_{i_2}\U_{i_3}^{T}\A\U_{i_4}\right\}\\
\doteq &L_1+L_2.
\end{align*}
Consider the first part $L_1$.
\begin{align*}
E&\left\{\sum_{2\le
j_1<j_2}^{n}\frac{1}{\tilde{n}_{j_1}^2(\tilde{n}_{j_1}-1)^2}\frac{1}{\tilde{n}_{j_2}^2(\tilde{n}_{j_2}-1)^2}\sum_{i_1,i_2=1}^{j_1-1}\sum_{i_3,i_4=1}^{j_2-1}
\U_{i_1}^{T}\A\U_{i_2}\U_{i_3}^{T}\A\U_{i_4}\right\}\\
=&E\left\{\sum_{2\le
j_1<j_2}^{n}\frac{1}{\tilde{n}_{j_1}^2(\tilde{n}_{j_1}-1)^2}\frac{1}{\tilde{n}_{j_2}^2(\tilde{n}_{j_2}-1)^2}\sum_{i=1}^{j_1-1}\sum_{i=1}^{j_2-1}
\U_{i}^{T}\A\U_{i}\U_{i}^{T}\A\U_{i}\right\}\\
&+E\left\{\sum_{2\le
j_1<j_2}^{n}\frac{1}{\tilde{n}_{j_1}^2(\tilde{n}_{j_1}-1)^2}\frac{1}{\tilde{n}_{j_2}^2(\tilde{n}_{j_2}-1)^2}\sum_{i_1=1}^{j_1-1}\sum_{i_2=1}^{j_2-1}
\U_{i_1}^{T}\A\U_{i_1}\U_{i_2}^{T}\A\U_{i_2}\right\}\\
&+E\left\{\sum_{2\le
j_1<j_2}^{n}\frac{1}{\tilde{n}_{j_1}^2(\tilde{n}_{j_1}-1)^2}\frac{1}{\tilde{n}_{j_2}^2(\tilde{n}_{j_2}-1)^2}\sum_{i_1=1}^{j_1-1}\sum_{i_2=1}^{j_2-1}
\U_{i_2}^{T}\A\U_{i_1}\U_{i_1}^{T}\A\U_{i_2}\right\}\\
\doteq &L_{11}+L_{12}+L_{13}.
\end{align*}
Taking the same procedure as Lemma 5 and some tedious calculations,
we can verify  that $L_{11}=o(\sigma_n^4)$,
$L_{12}+L_{13}=E^2(\eta_n)$ and $E(L_2^2)=o(\sigma_n^4)$. So,
$\var(\eta_n)=E(\eta_n^2)-E^2(\eta_n)=o(\sigma_n^4)$. This completes
the proof of (\ref{le3}).

Next, we proof result (\ref{lee4}). First of all, we note that
\begin{align*}
\sigma_n^{-2}\sum_{j=2}^{n}E[Z_{nj}^2I(|Z_{nj}|>
\epsilon\sigma_n|)|\mathcal{F}_{n,j-1}]\le
\sigma_n^{-4}\epsilon^{-2}\sum_{j=2}^{n}E[Z_{nj}^4|\mathcal{F}_{n,j-1}].
\end{align*}
Accordingly, the assertion of this lemma is true if we can show
\begin{align*}
E\left\{\sum_{j=2}^{n}E[Z_{nj}^4|\mathcal{F}_{n,j-1}]\right\}=o(\sigma_n^4).
\end{align*}
Notice that
\begin{align*}
E\left\{\sum_{j=2}^{n}E[Z_{nj}^4|\mathcal{F}_{n,j-1}]\right\}=\sum_{j=2}^{n}E(Z_{nj}^4)=O(n^{-8})\sum_{j=2}^{n}E\left(\sum_{i=1}^{j-1}\phi_{ij}\right)^4.
\end{align*}
Similar to \cite{r5}, the last term can be decomposed as $3Q+P$,
where
\begin{align*}
Q&=O(n^{-8})\sum_{j=2}^{n}\sum_{s\not=t}^{j-1}E(\U_j^{T}\U_s\U_s^{T}\U_j\U_j^{T}\U_t\U_t^{T}\U_j),\\
P&=O(n^{-8})\sum_{j=2}^{n}\sum_{s=1}^{j-1}E(\U_s^{T}\U_j)^4.
\end{align*}
Note that
\begin{align*}
Q=&O(n^{-8})\sum_{j=2}^{n}\sum_{s\not=t}^{j-1}E(\U_j^{T}\U_s\U_s^{T}\U_j\U_j^{T}\U_t\U_t^{T}\U_j)\}\\
=&O(n^{-8})\sum_{j=2}^{n}\sum_{s\not=t}^{j-1}E(\U_j^{T}\A\U_j\U_j^{T}\A\U_j)\\
=&O(n^{-1})\sigma_n^4,
\end{align*}
where the last equality is followed by
$E((\U_j^{T}\A\U_j)^2)=O(\tr(\A^2))$. Here we will show it.
\begin{align*}
\V_{1i}=&E(U(\Y_{1i}-\Y_{2j})|\Y_{1i})=E(U(\R^{1/2}(\bmv_{1i}-\bmv_{2j}))|\R^{1/2}\bmv_{1i})\\
=&E(U(\R^{1/2}(\bmv_{1i}-\bmv_{2j}))|\bmv_{1i})\\
=&E\left(\frac{\R^{1/2}(\bmv_{1i}-\bmv_{2j})}{||\R^{1/2}(\bmv_{1i}-\bmv_{2j})||}\bigg|\bmv_{1i}\right).
\end{align*}
Because of
$||(\R^{1/2}-\I_p)(\bmv_{1i}-\bmv_{2j})||^2=O(\tr(\R^{1/2}-\I_p)^2)=o(n^{-1}p^2)$,
then
\begin{align*}
||\R^{1/2}(\bmv_{1i}-\bmv_{2j})||=&||(\bmv_{1i}-\bmv_{2j})+(\R^{1/2}-\I_p)(\bmv_{1i}-\bmv_{2j})||\\
=&||\bmv_{1i}-\bmv_{2j}||(1+o_p(1)),
\end{align*}
and $\V_{1i}=\R^{1/2}\u_{1i}(1+o_p(1))$. Thus
\begin{align*}
E((\U_j^{T}\A\U_j)^2)=E((\u_{1i}^T\R^{1/2}\A\R^{1/2}\u_{1i})^2)(1+o(1))=O(\tr^2(\A^2)).
\end{align*}
Accordingly, we can verify that $Q=o(\sigma_n^4)$. In addition,
\begin{align*}
P=&O(n^{-8})\sum_{j=2}^{n}\sum_{s=1}^{j-1}E(\U_s^{T}\U_j)^4\\
=&O(n^{-8})\left\{\sum_{j=2}^{n_1}\sum_{s=1}^{j-1}E(\U_s^{T}\U_j)^4+\sum_{j=n_1+1}^{n}\sum_{s=1}^{n_1}E(\U_s^{T}\U_j)^4+\sum_{j=n_1+1}^{n}\sum_{s=n_1+1}^{j-1}E(\U_s^{T}\U_j)^4\right\}\\
\doteq &O(n^{-8})(P_1+P_2+P_3).
\end{align*}
As the procedures for handling $P_1,P_2,P_3$ are similar, let us
only consider $P_2$. By Lemma \ref{le1},
$E((\V_{1i}^T\V_{2j})^4)=O(\tr^2(\A^2)+\tr(\A^4))$, and then
$O(n^{-8}P_2)=o(\sigma_n^4)$. Similarly,
$O(n^{-8}P_1)=o(\sigma_n^4)$ and $O(n^{-8}P_3)=o(\sigma_n^4)$. This
completes the proof of (\ref{le4}). Thus, according to the
martingale central limit theorem \citep{r13}, we have
\begin{align*}
\frac{Z_n}{\sqrt{\var(Z_n)}}\cd N(0,1).
\end{align*}
Obviously,
\begin{align*}
\var(Z_n)=\frac{2}{n_1(n_1-1)}E((\V_{1i}^T\V_{1j})^2)+\frac{2}{n_2(n_2-1)}E((\V_{2i}^T\V_{2j})^2)+\frac{4}{n_1n_2}E((\V_{1i}^T\V_{2j})^2).
\end{align*}
So $\var(Z_n)=\sigma_n^2(1+o(1))$. Then we complete the proof of
this lemma. \hfill$\Box$